\documentclass[twocolumn,nofootinbib,prd,10pt]{revtex4}
\usepackage{graphicx}
\begin{document}

\title{Orthographic Correlations in Astrophysics}
\author{Joe Zuntz}
\affiliation{Astrophysics Group, University of Oxford, United Kingdom}
\author{Thomas G. Zlosnik}
\affiliation{Perimeter Institute of Theoretical Physics, Waterloo, Canada}
\author{Caroline Zunckel}
\affiliation{Astrophysics Department, Princeton University, New Jersey, USA}
\affiliation{Astrophysics and Cosmology Research Department, University of KwaZulu-Natal, South Africa}
\author{Jonathan T.~L. Zwart}
\affiliation{Columbia Astrophysics Laboratory, Columbia University, 550 West 120th Street, NY 10027, USA}
\date{April 1st 2010}

\begin{abstract}
We analyze correlations between the first letter of the name of an author and the number of citations their papers receive.  We look at simple mean counts, numbers of highly-cited papers, and normalized h-indices, by letter.  To our surprise, we conclude that orthographically senior authors produce a better body of work than their colleagues, despite some evidence of discrimination against them.
\end{abstract}
\maketitle

\section{Introduction}
Citation counts and indices based on them are a key aspect of the sociology and practice of astrophysics, and of many other fields.  Various metrics to associate the citation counts of an author with their value as a scientist have been developed, such as the h-index, h-b-index, and g-index\cite{indices}, though all have been criticized as inappropriate in different contexts.  Indeed, any metric based solely on citation counts is ultimately flawed, since it is unable to distinguish between citations of the form ``Work in \cite{zunckel1} forms the basis of this discussion'' and ``We find the results in \cite{zuntz1} to be utterly incorrect''.  Some unscrupulous authors even inappropriately cite their own work to improve their indices.

Despite their flaws, citation counts and indices certainly do have an important impact, both on science and scientists.  Highly cited papers become part of the canon of standard works, and the authors of highly-cited papers are employed and feted\footnote{An earlier version of this manuscript misspelled this word as \emph{fetid}; we apologize for any confusion caused.}.  Characterizations of patterns of citation can therefore be both metrics for wider socio-scientific trends and indicators of surprising aspects of paper generation, publication and promulgation systems.

There have been previous studies of the correlation between number of citations and the length \cite{Stanek1}, number of authors \cite{Stanek2}, release timing \cite{dietrich,haque}, field \cite{leydesdorff} and publication method \cite{gentil} of articles.  Here we study another aspect of citation patterns: correlation between the number of citations received by an article and its orthography.  Specifically, we correlate the number of citations a paper has with the first letter of its first author's surname.  Selection by paper rather than by author does pose some problems, but it is significantly more tractable to collect such data given inconsistency in citation names.

In Section \ref{hypothesis and data section} we discuss the hypotheses we wish to test and the data used.  In Section \ref{analysis section} we analyze our data and discuss various citation measures.  We conclude in Section \ref{sec:discussion}.

\section{Hypotheses \& Data\label{hypothesis and data section}}
We can immediately lay out a near-exhaustive hypothesis space of this problem:
\begin{description}
\item[H0] Authors with names near the end of the alphabet (AWNNTEOTAs) have more citations, because of some intrinsic superiority\footnote{We use the word \emph{superiority} here somewhat tautologously, simply to indicate the number of citations received.  We do not intend to imply any superiority in ability, intellect or physical appearance.}.
\item[H1] AWNNTEOTAs have fewer citations, because they are discriminated against.
\end{description}
We have omitted a third possibility, that there is no significant difference in the citation counts, owing to its evident implausibility.

To test these hypotheses we use data from the SAO/NASA Astrophysics Data System (ADS).  We use data from 12 randomly selected months, extracted using that site's query tool: we list all the astrophysics papers published in that month.  We extract the number of citations to each article whenever that information is available.  The months selected  are listed in Table \ref{months table}.  
\begin{table}
\caption{Months analyzed.}
\begin{center}
\begin{tabular}{|c|c|c|c|}
\hline
Month & Year & Month & Year \\
\hline
June & 2000 & May & 2006 \\
November & 2000 & August & 2006 \\
December & 2001 & February & 2007 \\
April & 2002 & May & 2007\\
August & 2002 & November & 2007\\
January & 2005 & February & 2009\\
\hline
\end{tabular}
\end{center}
\label{months table}
\end{table}

We reduce the data such that for each letter of the alphabet we have a list of all the citation counts of papers written by authors whose name starts with that letter.  For some months the  more exclusive letters have no first-author publications.  The months cover a long enough period that both new papers and more highly cited papers are included.

\section{Analysis\label{analysis section}}
In this section we consider how various citation measures correlate with alphabetic position.  We need to be careful here to tease out effects that arise from correlation between position and popularity of a letter; rigorous statistical techniques are imperative.  To efficiently and understandably analyze our data we adopt the usual astrophysical paradigm: \emph{Bayesio-frequentist} statistics, where frequentist methods are used and the results interpreted as though they were Bayesian probabilities. We also follow usual practice and use a number of different estimators, continuing until we find one that can demonstrate the correct hypothesis to be true.

\subsection{Mean citation counts}

\begin{figure}[htbp]
\begin{center}
\includegraphics[width=\columnwidth]{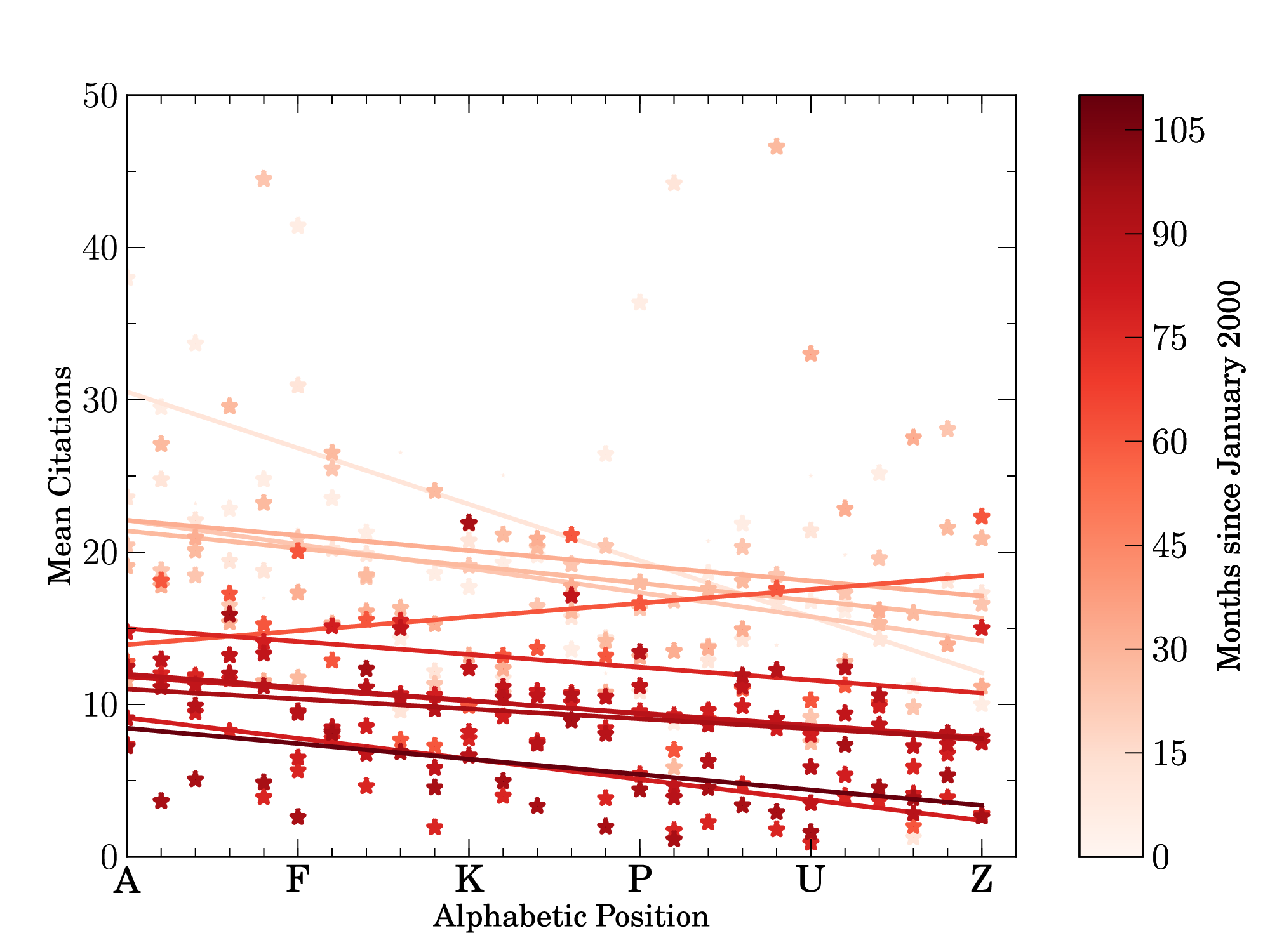}
\caption{Mean citation counts with alphabetic position and time.  Horizontal location indicates alphabetic position and the color indicates time.  The stars are individual data points and the solid lines are linear trends for each month.}
\label{mean plot}
\end{center}
\end{figure}

The simplest metric we can use is just the mean citation count for a letter.  These values are plotted in Figure \ref{mean plot}, and the trend-line coefficients shown in Table \ref{means table}.   The general trend of decreasing citations with time simply indicates the lack of time for recent papers to accumulate citations.  Disturbingly, these data seem to prefer hypothesis H1, since they show a decrease in mean citations counts with alphabetic position, nearly consistently across the months surveyed.

\begin{table}
\caption{Linear fit coefficients for $c = a x + b$ where $c$ is the mean number of citations for a letter with alphabetic position $x$.  Only one month, January 2005, shows the expected trend of more citations for orthographically advanced authors.}
\begin{center}

\begin{tabular}{|c|c|c|}
\hline
Month & $a$ & $b$ \\
\hline
06/00 & $-0.033$ & $21.322$\\
11/00 & $-0.738$ & $30.523$\\
12/01 & $-0.317$ & $22.096$\\
04/02 & $-0.229$ & $21.395$\\
08/02 & $-0.199$ & $22.093$\\
01/05 & $0.181$ & $13.925$\\
05/06 & $-0.169$ & $14.975$\\
08/06 & $-0.270$ & $9.127$\\
02/07 & $-0.159$ & $11.818$\\
05/07 & $-0.173$ & $11.993$\\
11/07 & $-0.130$ & $11.008$\\
02/09 & $-0.203$ & $8.442$\\
\hline
\end{tabular}
\end{center}
\label{means table}
\end{table}

\begin{figure}[htb]
\begin{center}
\includegraphics[width=\columnwidth]{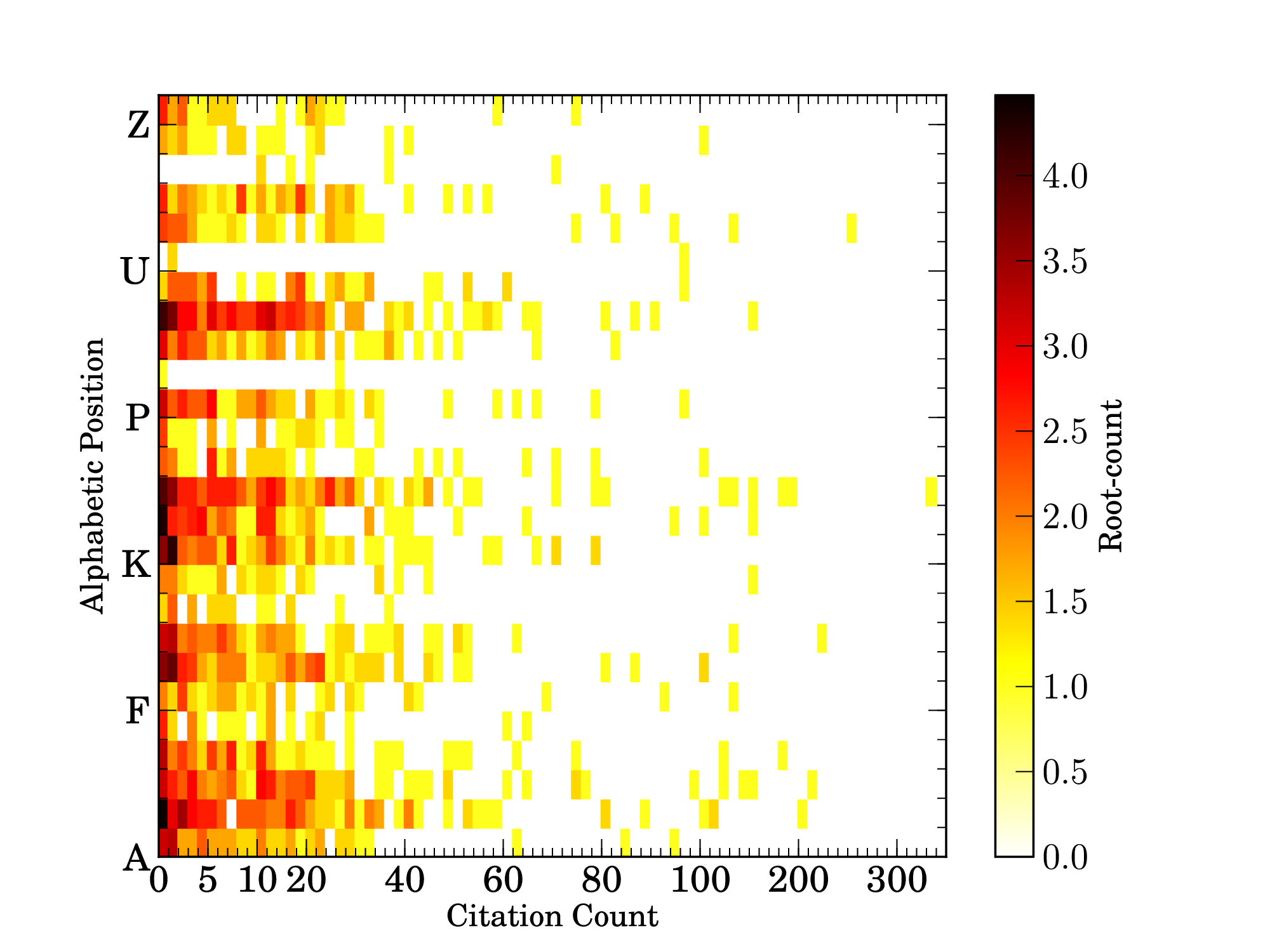}
\caption{Citation statistics for January 2005.  The citation count axis is scaled and binned to aid comprehension; the axis progresses linearly between each tick mark. Color indicates the square-root of the number of papers in the bin for this month.}
\label{hotplot}
\end{center}
\end{figure}

As noted in Table \ref{mean plot}, only one month shows a positive trend with alphabetic seniority.  We show statistics from this month, January 2005, in more detail in Figure \ref{hotplot}.  Inspection of the plot suggests that the different statistical behavior that month was largely due to an impressive performance by the letter ``S'' and below-par performances by ``A'' and ``B''.

\subsection{Highly Cited Papers}
Unlike real doctors, scientists are forgiven their worst work and judged on their best.  The number of uncited papers\footnote{Paradox prevents us citing such a paper here.} produced by an author is irrelevant if they have any highly cited ones.  We therefore consider the most highly cited papers over all the data.  We choose the top 5\% of papers as our benchmark for ``highly cited'', since we need a large enough number to get reasonable statistics.  This typically means dozens of citations (for the very recent months) to hundreds or thousands of citations (for the oldest months in our set).  We must also normalize by the total number of authors with a given initial letter to account for the rarity of certain letters.

\begin{figure}[htb]
\begin{center}
\includegraphics[width=\columnwidth]{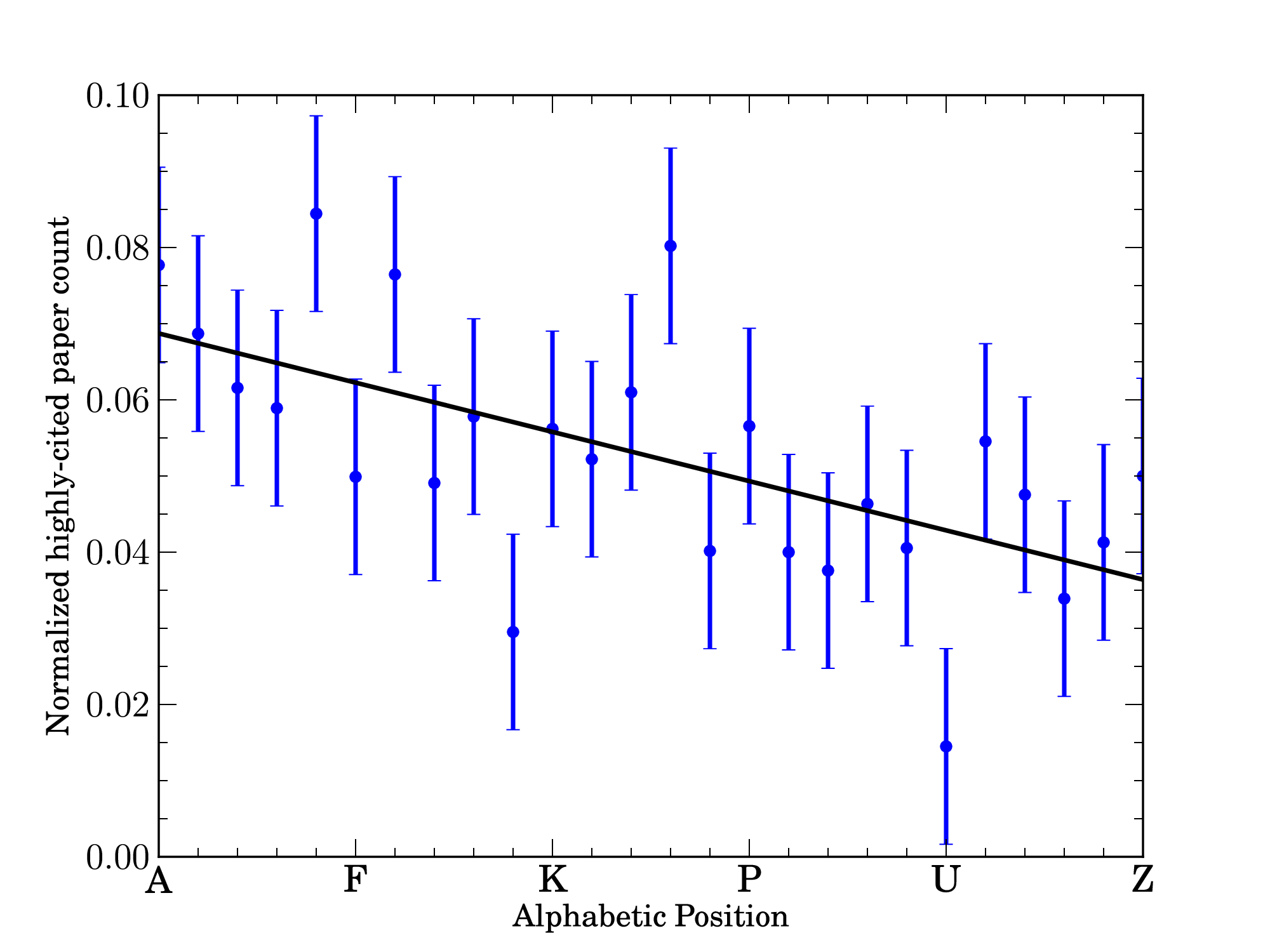}
\caption{The number of highly cited papers produced with initial author letter, together with a trend line.  Point errors are computed using a procedure adapted from supernova cosmology, where the errors are altered until the reduced $\chi^2\sim1$.}
\label{highly cited plot}
\end{center}
\end{figure}

The results are shown in Figure \ref{highly cited plot}, and they show a clear trend downwards with the progression of the alphabet, once again supporting hypothesis H1.  We therefore try another analysis method.

\subsection{Normalized h-indices}
\begin{figure}[htb]
\begin{center}
\includegraphics[width=\columnwidth]{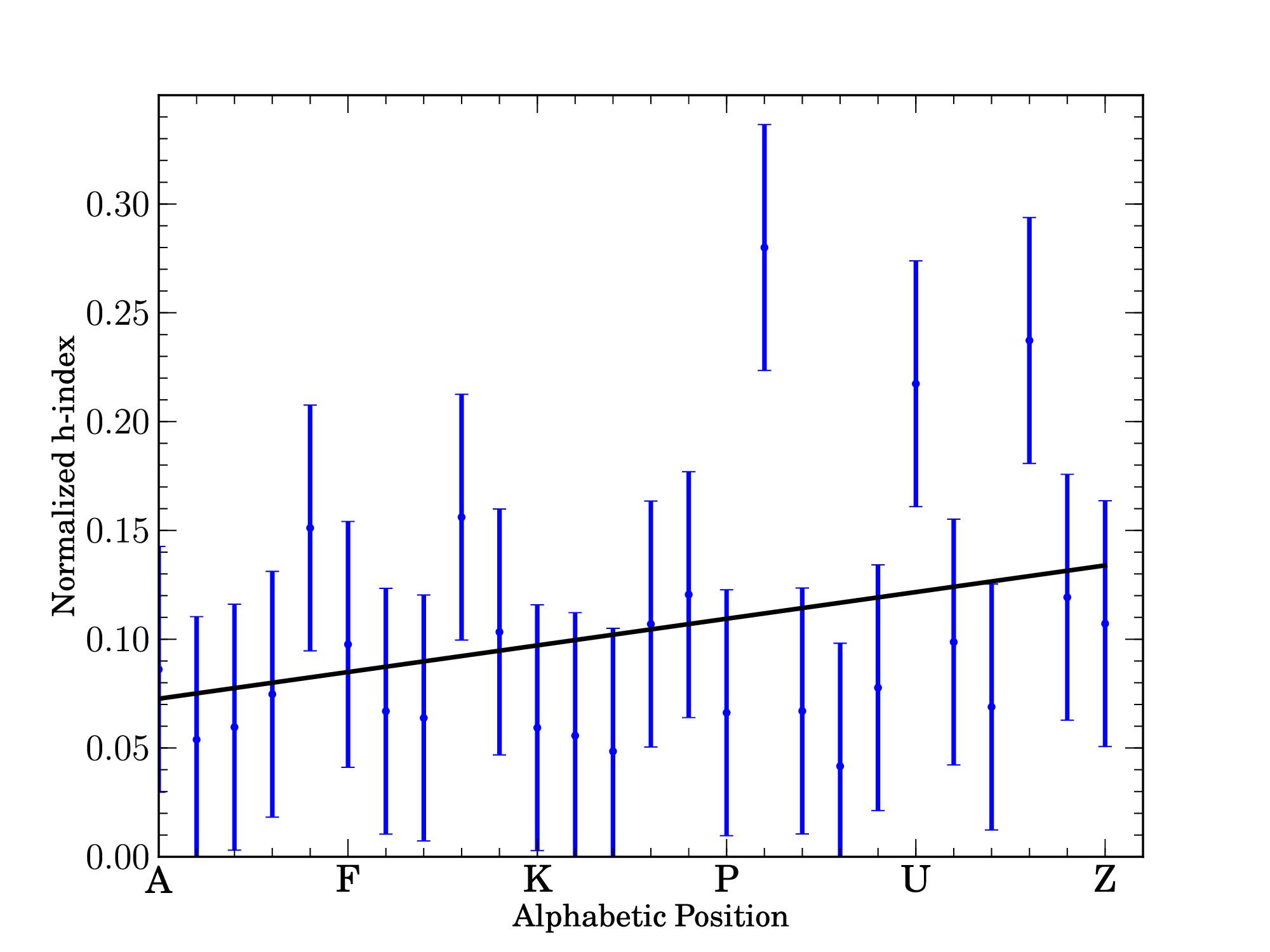}
\caption{The normalized h-index for our global data set with alphabetic position (blue points) and a trend line (black solid).  Point errors are produced as for Figure \ref{highly cited plot}.}
\label{normalized h index plot}
\end{center}
\end{figure}

The standard h-index cannot be applied to the collected data, since it does not take into account the number of authors with a given initial letter.  We therefore normalize it by the number of authors in the full data.  The result for the collated complete data set, with all the months collected together, is shown in Figure \ref{normalized h index plot}.

As shown by the trend line, the result clearly supports hypothesis H0; AWNNTEOTAs have a higher normalized h-index.  Furthermore, the same effect appears whether we look at the bulk of the data points or consider only the extremal values in the former or latter halves of the data.

\section{\label{sec:discussion}Discussion}
Based on the results generated, our third method of h-indices appears to be the most reliable; we are therefore reluctantly directed towards hypothesis H0, that orthographically high-ranking authors produce a better global body of work than their less alphabetically gifted colleagues.

Two caveats apply to our selection of data.  First, we are somewhat restrictive by limiting ourself to first authors only, particularly in light of the practice of alphabetizing authors on some papers, though this is more common for authors after the first.  It is possible that any discrimination against or superiority of AWNNTEOTAs occurs at the point of choosing which author should be first. Second, we use the data themselves to estimate the raw number of authors with a given initial letter.  This might skew results, if, for example, there are large numbers of authors with the surname ``Aaaronson'' who produced no papers at all during the referenced period; this would make our conclusions stronger.

While future work on this topic could possibly improve on the analysis and interpretation methods we have used here, our conclusions can be made significantly more robust by improving the data directly \cite{zwart}\cite{zuntz2}\cite{zuntz3}\cite{zlosnik1}\cite{zwolak}\cite{zlosnik2}\cite{zlosnik3}\cite{zhang}\cite{zunckel2}\cite{zunckel3}\cite{zurek}\cite{zunckel4}.
 
\emph{Acknowledgments.}
We would like to acknowledge the help of various colleagues in formulating this work, but they asked us not to because they didn't want their names associated with it.  The exception was Laura Newburgh, who suggested our method for improving future data.

\end{document}